# 3D-2D Neural Nets for Phase Retrieval in Noisy Interferometric Imaging

Andrew H. Proppe[1,2,3], Guillaume Thekkadath[2], Duncan England[2], Philip J. Bustard[2], Frédéric Bouchard[2], Jeff S. Lundeen[1,3,*], Benjamin J. Sussman[1,2,*]

[1] Joint Center for Extreme Photonics, University of Ottawa and National Research Council of Canada, 100 Sussex Drive, Ottawa, Ontario, Canada K1A 0R6
[2] National Research Council of Canada, 100 Sussex Drive, Ottawa, Ontario K1A 0R6, Canada
[3] Department of Physics and Nexus for Quantum Technologies, University of Ottawa, 25 Templeton Street, Ottawa, Ontario, Canada K1N 6N5

**Author email addresses:**
aproppe@uottawa.ca; guillaume.thekkadath@nrc-cnrc.gc.ca; Duncan.England@nrc-cnrc.gc.ca; philip.bustard@nrc-cnrc.gc.ca; frederic.bouchard@nrc-cnrc.gc.ca; jlundeen@uottawa.ca, bsussman@uottawa.ca

**Corresponding author contact details:**
Professor Jeff Lundeen
University of Ottawa
25 Templeton St, Ottawa, ON K1N 6N5
jeff.lundeen@gmail.com, (613) 562-5800 X 7637

Professor Benjamin Sussman
National Research Council of Canada
100 Sussex Drive, Ottawa, Ontario, Canada K1N 5A2
bsussman@uottawa.ca, (613) 991-6888

**Abstract**

**In recent years, neural networks have been used to solve phase retrieval problems in imaging with superior accuracy and speed than traditional techniques, especially in the presence of noise. However, in the context of interferometric imaging, phase noise has been largely unaddressed by existing neural network architectures. Such noise arises naturally in an interferometer due to mechanical instabilities or atmospheric turbulence, limiting measurement acquisition times and posing a challenge in scenarios with limited light intensity, such as remote sensing. Here, we introduce a 3D-2D Phase Retrieval U-Net (PRUNe) that takes noisy and randomly phase-shifted interferograms as inputs, and outputs a single 2D phase image. A 3D downsampling convolutional encoder captures correlations within and between frames to produce a 2D latent space, which is upsampled by a 2D decoder into a phase image. We test our model against a state-of-the-art singular value decomposition algorithm and find PRUNe reconstructions consistently show more accurate and smooth reconstructions, with a ×2.5 - 4 lower mean squared error at multiple signal-to-noise ratios for interferograms with low (< 1 photon/pixel) and high (~100 photons/pixel) signal intensity. Our model presents a faster and more accurate approach to perform phase retrieval in extremely low light intensity interferometry in presence of phase noise, and will find application in other multi-frame noisy imaging techniques.**

## Introduction

Phase retrieval is a fundamental problem in physics that underpins many imaging technologies, including quantitative phase microscopy, holography, optical coherence tomography, coherent diffractive imaging, wavefront sensing, and crystallography. The problem involves the reconstruction of a wave's phase distribution from intensity-only measurements of its diffraction or interference patterns. This reconstruction is generally highly sensitive to noise in the data, and thus has traditionally relied on regularization techniques and optimization algorithms like maximum likelihood estimation. But in recent years, there has been a paradigm shift in addressing the phase retrieval problem due to the emergence of deep convolutional neural networks (NNs).[1,2] By training NNs on datasets containing structures relevant to the application of interest, the regularization process becomes tailored towards reconstructing those structures. As a result, NNs can solve the phase retrieval problem with superior accuracy and speed compared to conventional optimization algorithms, especially in the presence of noisy data.[3–11]

In interferometry, noise in the measured interference patterns, or interferograms, can arise from diverse sources. These include (i) shot noise resulting from fundamental quantum intensity fluctuations in the light fields, (ii) detector noise encompassing both read and thermal noise, and (iii) phase noise due to vibrations or other instabilities in an interferometer. While previous NNs architectures have demonstrated phase retrieval in presence of the first two types of noise, even with interferograms containing only a few photons per detector pixel,[4,10,11] the third source of noise presents a different challenge. Unlike detector and shot noise, phase noise cannot be mitigated by extending the exposure time of the interferograms, as doing so would "wash-away" the interference pattern containing the phase information. For example, in remote sensing scenarios, atmospherically-induced phase fluctuations and vibrations restrict exposure times to sub-millisecond durations.[12] Short exposure times in turn limit the signal-to-noise ratio (SNR) of the interferograms, especially in applications where the illuminating power is constrained, such as with photosensitive samples[13] or when the object of interest is poorly reflective.

Recently, it was demonstrated that phase noise can be overcome by temporally-resolving the interferograms and measuring a time series of randomly phase-shifted "frames".[14–16] While the exposure time of the individual frames must be shorter than the stability time of the interferometer, the time series can be accumulated for arbitrary durations. Due to the randomly varying phase offsets between frames, the challenge is to retrieve the object phase from correlations contained across all frames. So far, this challenge has been addressed using a principal component analysis[15,17] and Fourier filtering[14,15] on the second-order correlation function of the time series. Unfortunately these techniques are insensitive to higher-order correlations which also contain information about the object phase[18]. While NN are naturally suited for this task due to their multi-layer convolutions, current NN architectures for phase retrieval only allow for a handful of interferograms as an input.[3–11] A detailed account of recent works related to ours is given in the Supplementary Information (SI).

In this work, we overcome the challenge of randomly varying phase offsets using a deep convolutional NN that performs the phase retrieval on a 3D ensemble of noisy and randomly phase-shifted intensity interferograms to produce a single 2D phase output. In our experiments, we place a 2D phase object, $\Phi(r)$, in one arm of a balanced optical interferometer, which modulates the phase but not the amplitude of the passing light beam (Fig. 1a and b). Here, our phase object is a spatial light modulator, which allows us to program 2D patterns. The phase-modulated beam is recombined with the reference beam on a 50:50 beamsplitter, giving a nominal interference visibility of 100% across the image, and the resulting series of interferograms are detected on a 2D camera sensor.

For an interferometer with long path lengths, random phase modulations due to vibrational instability, air currents (turbulence), and thermal drift will occur before the beams recombine and travel to the camera. Here, we consider only a weak turbulence regime, where the phase fluctuations are uniform across the image and do not have a spatial (pixelwise) dependence. This random phase-shifting is clearly observed in Fig. 1a, where the true $\Phi(r)$ is shown alongside four intensity images collected by the experimental setup. Each frame or shot $i$ contains the same spatial structure; but the random phase fluctuations, $\varphi_i$ (which here are spatially uniform) can completely randomize the relative shot-to-shot intensities of the object.

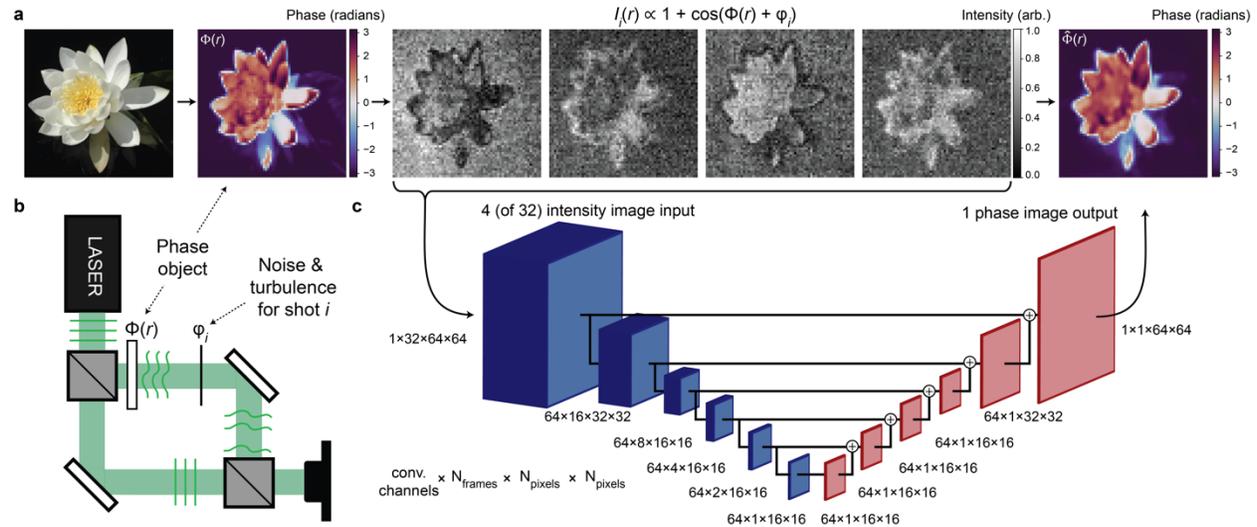

**Figure 1.** Problem formulation, optical setup, and neural network architecture. **(a)** A colored image is converted into a 2D phase image, $\Phi(r)$, from which several frames are measured with a limited acquisition time. Each frame accumulates random Gaussian and Poisson noise, and the frames are also randomly phase-modulated by noise and turbulence sources that vary shot-to-shot, $\varphi_i$. Here, $\varphi_i$ is a uniform random phase offset between 0 and $2\pi$. **(b)** Schematic of an optical setup where random, spatially uniform phase modulation is encountered, with $\varphi_i$ fluctuating randomly on a ~10 ms timescale. **(c)** Schematic architecture of our **P**hase **R**etrieval **U**-**Ne**t (PRUNe) that converts a 3D ensemble of randomly phase-modulated intensity interferometry images into a single 2D image of the phase object, $\hat{\Phi}(r)$. Detailed diagrams of the downsampling convolutional layers (blue) and upsampling convolutional transpose

layers (red) are given in Fig. S1. The + symbols indicate skip connections that combine encoder outputs with decoder inputs by elementwise addition.

To address this challenge, we employ a neural network model that makes use of 3D convolutional layers to encode the ensemble of randomly phase-modulated intensity images, and 2D convolutional layers to decode a single 2D output of the phase object. The 3D convolutional layers capture correlations within and between neighbouring frames: for example, it learns the spatial structural motifs of our phase objects, and learns them within the context of their relative intensities being randomly shifted frame-to-frame. We benchmark the performance of our model against a singular value decomposition (SVD) algorithm which performs a principal component analysis of the second-order correlation function averaged over all frames.[8,16] This algorithm is the current state-of-the-art for phase retrieval from randomly phase shifted frames, surpassing the performance of alternatives like the advanced iterative algorithm.[19]

**Neural network architecture, data, and training**

Deep convolutional networks have been used extensively for computer vision tasks like image recognition,[20] segmentation,[21] and restoration or denoising.[22] A recent review of image denoising with different deep learning approaches is given by Izadi et al. in ref. [23]. One such approach is to use an autoencoder architecture,[24] which are well explored for denoising tasks on inputs with normally distributed (Gaussian) noise,[25] as well as for shot noise in the case of low-intensity data with few counts.[26,27]

Our network architecture, shown in Fig. 1c, has the general structure of a convolutional autoencoder: the encoder path uses 3D convolutional operations, separately controlling the downsampling for the spatial and frame dimensions at each block, until the frame dimension is reduced to 1. In this way, we can minimize unnecessary compression of the image's spatial dimensions while still reducing our frame dimension. The decoder then upsamples from the latent space using 2D convolutional transpose operations until the original spatial dimensions of a single 2D image are restored. Each layer of our encoder (decoder) is a residual-network-type block,[20] containing two convolutional (convolutional transpose) layers, each followed by a batch normalization and nonlinear activation layer, and with an optional skip connection established between the input and output. See Fig. S1 for a diagram and details of each block.

Inspired by U-Net architectures,[21] the encoder and decoder paths are symmetric (with respect to the spatial dimensions) and have skip connections between the output of each encoder downsampling layer and its corresponding decoder upsampling layer. This overcomes gradient vanishing and allows for deeper networks.[20,22] We thereby named our network **P**hase **R**etrieval **U**-**Ne**t: PRUNe. The encoder residual outputs are 3D, whereas the decoder inputs are 2D, and so to allow symmetric skip connections, we simply averaged over the frame dimension of the encoder output before adding it to the decoder input.

Our model training uses 5000 images of flowers taken from the dataset given in ref. 28. We selected this dataset as a test case to demonstrate phase retrieval on images of natural and realistic objects, which have greater spatial variety, more complex phase gradients, and features on multiple size scales, in contrast to e.g. MNIST handwritten digits. The same sets of images were used for both experimentally measured and simulated interferograms. Each time a set of interferograms was loaded as training data, the order of the frames was randomly shuffled, to ensure uncorrelated phase jumps between frames and to help prevent overfitting. We used three different losses between the ground truth images and the PRUNe reconstructions: mean squared error (MSE), structural similarity (SSIM),[29] and gradient difference loss (GDL). The MSE loss ensures pixel-by-pixel accuracy of the reconstructions, whereas the SSIM and GDL loss functions are used to preserve finer image details and ensure sharper reconstructions. We found that incorporating the SSIM and GDL losses led to faster learning and lower MSE in the reconstructions than models trained only using MSE; and that models trained using only SSIM and GDL without MSE produced poorer results, indicating the synergistic rather than individualistic utility of these loss functions for our problem (Fig. S2). Optimal hyperparameters for our PRUNe model were obtained by sweeping 50 different combinations using the package Weights & Biases,[30] and are tabulated in Table S1 and Fig. S3. Training and validation losses versus number of steps (epochs) for optimized models are shown in Fig. S4. Code and values for the data generation, training, and evaluation pipeline can be found in the GitHub repository.

## Results

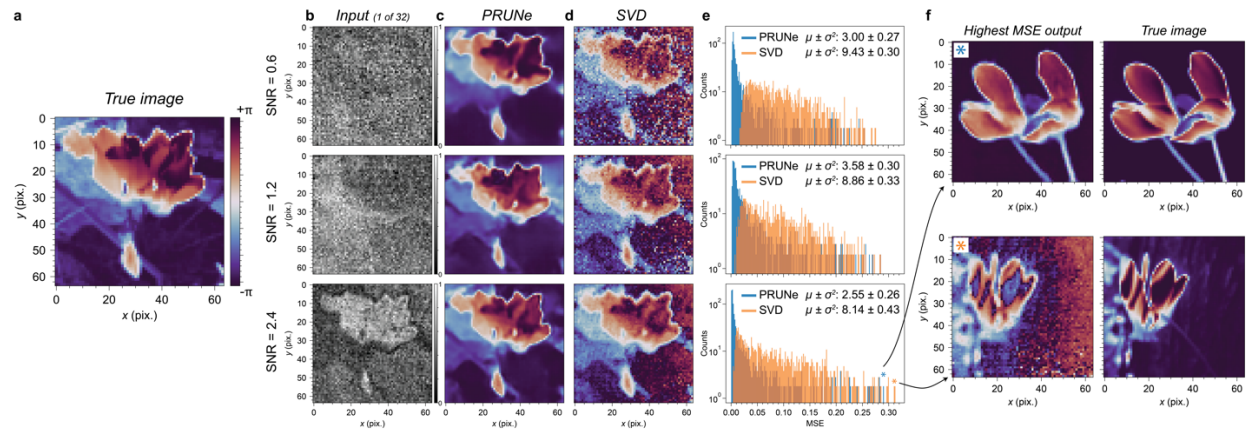

**Figure 2.** Results for experimental high signal, high background data. **(a)** True phase image, **(b)** 1 of 32 input frames, **(c)** PRUNe reconstruction, **(d)** SVD reconstruction, and **(e)** histogram of MSE for 1000 different images. Rows from top to bottom correspond to different signal-to-noise ratios (SNR) for the input frames, obtained by changing the acquisition time as 0.025, 0.050, and 0.100 ms per frame. The true image and reconstructions shown here correspond to the lowest MSE reconstruction by PRUNe at the highest SNR. The PRUNe and SVD reconstructions with the highest MSE and corresponding true phase images are shown in panel **(f)**.

To explore the versatility of our neural network approach for different levels of signal intensity, we consider two different measurement regimes, which are classified based of the number of signal photons per pixel, $\bar{n}_s$, and the number of background photons per pixel, $\bar{n}_b$, summarized in Table 1. The high signal regime of $\bar{n}_s$ = 50 - 200 emerges in more routine imaging measurements using standard cameras with high background counts ($\bar{n}_b$) and laser pulses containing a large number of photons. The low signal regime of $\bar{n}_s$ < 1 could occur using faint light sources paired with a single-photon sensitive camera, which may be important for applications involving photosensitive samples.[9,13] There are important distinctions between the two regimes: for high $\bar{n}_s$, the model must learn to distinguish between the signal and background counts, and the interferograms will be dominated by large detector read-out noise. For low $\bar{n}_s$, since the object is undersampled, the model must learn to infer the structure from only a few photons, and shot noise will dominate.

Within both regimes, we consider three different SNRs. For the high signal regime, we use experimental data, collected using the optical setup described in Fig. 1, where the SNR is varied by changing the acquisition time of each frame. For the low signal regime – which was not experimentally feasible owing to the large readout noise of our camera – we instead use simulated data, where series of interferograms are generated with fixed $\bar{n}_s$ and variable $\bar{n}_b$ (values in Table 1) to change the SNR over three orders of magnitude. Details of generating simulated image frames are given in the Methods and visualized in the SI. Two different PRUNe models were separately trained: one for high signal regime using experimental data, and for the low signal regime using simulated data.

| Signal Level | $\bar{n}_s$ (photon/pixel) | $\bar{n}_b$ (photon/pixel) | SNR | MSE $\mu \pm \sigma^2$ (PRUNe, ×10³) | MSE $\mu \pm \sigma^2$ (SVD, ×10³) | $\mu$ ratio (SVD/PRUNe) |
|---|---|---|---|---|---|---|
| High | 53 | 88 | 0.6 | 3.00 ± 0.27 | 9.43 ± 0.30 | 3.14 |
| | 106 | 88 | 1.2 | 3.58 ± 0.30 | 8.86 ± 0.33 | 2.47 |
| | 212 | 88 | 2.4 | 2.55 ± 0.26 | 8.14 ± 0.43 | 3.19 |
| Low | 0.25 | 0.25 | 1 | 1.85 ± 0.09 | 7.89 ± 0.14 | 4.26 |
| | 0.25 | 0.025 | 10 | 1.83 ± 0.10 | 5.59 ± 0.13 | 3.05 |
| | 0.25 | 0.0025 | 100 | 1.81 ± 0.20 | 5.28 ± 0.13 | 2.91 |

**Table 1.** The average number of photons per pixel measured from signal from the phase target ($\bar{n}_s$) and from background counts ($\bar{n}_b$), for two different regimes of high versus low signal. Means ($\mu$) and variances ($\sigma^2$) of the mean-squared error (MSE) for datasets of 1000 images analyzed by our PRUNe model and the SVD algorithm. MSEs are scaled by $10^3$ for clarity.

**High signal regime**

Figure 2 shows results for PRUNe (trained on experimental data) and SVD reconstructions in the high signal regime using experimentally measured inputs consisting of 32 frames of 64×64 pixels for the three different SNRs, quantified in Table 1. The acquisition time for each of the frames was 0.025, 0.050, and 0.100 ms, respectively, for the three different SNRs. To compare the statistical error of our models, we evaluated the MSE of 1000 different reconstructions, which compose the validation dataset used while training the neural network models. The true image (Fig. 2a), experimental inputs (Fig. 2b), and reconstructions by PRUNe (Fig. 2c) and SVD (Fig. 2d) shown here correspond to the PRUNe reconstruction with the lowest MSE out of the set of 1000 for the highest SNR. Several more reconstructions by both PRUNe and SVD are shown in Fig. S5, and the full arrays of 32 data inputs in Fig. S6.

We observe that both models fail to capture some of the finer details of the true image: in particular the veins and stems of the some of the background leaves. We observe that PRUNe produces much smoother overall reconstructions and correctly captures the phase across the entire image. In comparison, SVD results in noise across the entire frame, and while it captures the salient features of the centered flower object, it returns incorrect flat phases in the surroundings, particularly around the righthand side of this image - and interestingly, this is true for all SNRs, despite an overall reduction in noise.

A histogram of the MSEs for the PRUNe and SVD methods are shown in Fig. 2e, along with the mean ($\mu$) and variance ($\sigma^2$) of the MSE distributions, which are summarized, along with the ratios of $\mu$ for PRUNe and SVD, in Table 1. Both models trend to lower error as the SNR increases (note the increasing scale on the counts axis). The minimum error for the SVD reconstructions trends lower with increasing SNR, whereas this is not observed for PRUNe. For all SNRs, our PRUNe model exhibits lower average MSE and lower variance in the MSE compared to SVD, confirming the superior phase retrieval capabilities of our neural network model across the 1000 test images.

To investigate the limitations and weaknesses of each model, we also show their reconstructions with the highest MSEs in the highest SNR dataset, shown in Fig. 2f. The PRUNe reconstruction exhibits mostly the same characteristics as the best case: loss of some finer details, but an overall smooth image with correct phase across the entire frame. In contrast, the SVD reconstruction appears defocused, shows random pixelated noise, and completely misses the correct phase on the righthand side of the image that is mostly flat. This is consistent with our hypothesis that SVD performs poorly on flat phase distributions where one of the principal components can vanish;[16,17] a limitation that does not affect our PRUNe model.

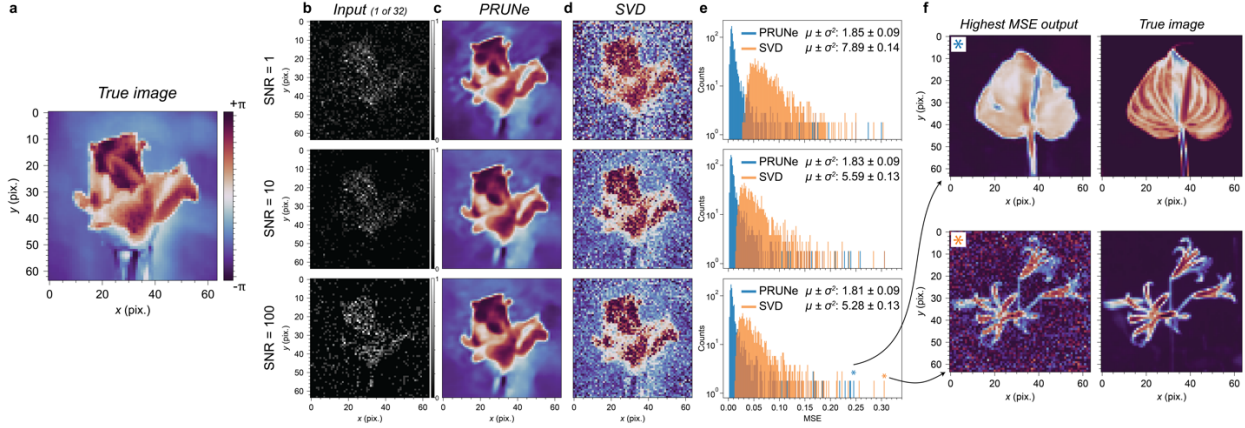

**Figure 3.** Results for experimental low signal, low background data. **(a)** True phase image, **(b)** 1 of 32 input frames, **(c)** PRUNe reconstructions, **(d)** SVD reconstructions, and **(e)** histograms of MSE for 1000 different images. Rows from top to bottom correspond to different signal-to-noise ratios (SNR) for the input frames, obtained by changing the average number of simulated signal and background photons per pixel (Table 1). The true image and reconstructions shown here correspond to the lowest MSE reconstruction by PRUNe at the highest SNR. The PRUNe and SVD reconstructions with the highest MSE and corresponding true phase images are shown in panel **(f)**.

**Low signal regime**

We turn now to the low signal regime of < 1 photon/pixel in Figure 3, using simulated rather than experimental inputs again consisting of 32 frames of 64×64 pixels for the three different SNRs, quantified in Table 1. As with the high signal regime, we show here the true image (Fig. 3a), simulated inputs (Fig. 3b), and reconstructions by PRUNe (Fig. 3c) and SVD (Fig. 3d) corresponding to the PRUNe reconstruction with the lowest MSE out of the set of 1000 for the highest SNR. Fig. S7 shows more reconstructions in the low signal regime, and Fig. S8 shows examples of the full 32 input frames. Whereas PRUNe affords smooth reconstructions across all three SNRs, the SVD consistently exhibits large amounts of pixelated noise. Both models again do not recover some of the finer details.

Similar to the high signal regime, PRUNe outperforms the SVD algorithm at all SNRs. However, while the MSE decreases with decreasing background for SVD, it stays the same for PRUNe, suggesting that the model is more robust to background noise in the low-signal regime. We observe an even greater relative performance by PRUNe over SVD, indicated by the larger ratios of $\mu$ shown in Table 1: a ×4.26 improvement at the lowest SNR.

We again show the worst (highest MSE) reconstructions in the highest SNR dataset for both models in Fig. 3f. Here, we observe rather poor performance by both models. PRUNe obtains an overall correct phase for the object and background, but completely fails to capture finer details of the leaf pattern. The SVD performs well in capturing the foreground object, but shows completely randomized phase for the flat background; this latter point is consistent with our observations in the high signal regime.

**Discussion**

Our interferometric phase retrieval problem required a neural network model that not only removes random noise from each of the frames, but additionally accounts for the random frame-to-frame phase-shifting within the ensemble. As opposed to 2D convolutional networks that process individual images, 3D convolutional operations can instead be used for a series of images, like video[31] or multiple cross-sections from medical scans.[32] In such cases, the 3D convolutional kernels provide the advantage of being able to capture local correlations across frames, as well as spatial dimensions within each frame. Our problem presented an interesting contrast to previous investigations using 3D convolutional neural networks: rather than a set of images that evolve deterministically in time (e.g. a video) or space (e.g. 3D medical scans), we instead have an ensemble of images with random phase jumps between frames that affect the entire imaged plane. We reasoned that, compared to a purely 2D convolutional autoencoder that operates frame-by-frame, 3D convolutional operations would still be advantageous in our case to capture spatial correlations within and between different frames.

We also adopted a U-Net architecture, since it is of the most performant modern variants of convolutional networks.[22] U-Net-like architectures using 3D convolutional layers have been previously reported, particularly for the field of medical imaging.[33–35] In our case however, each of our inputs represent the same underlying noise-free phase image, and we convert a 3D ensemble input into a single 2D output. To the best of our knowledge, this is one of the first implementations of dimensionality-reducing 3D-to-2D (simplified to 3D-2D) U-Net, where an image dimension is gradually removed within the U-Net itself, rather than in reshaping layers before or after.

This 3D-2D U-Net strategy provides a model that exhibits superior performance compared to the SVD algorithm for two different overall light intensity regimes, and at multiple SNRs in each regime. From our experiments, we find that PRUNe affords smoother reconstructions with lower MSE and correct phase across the entire image frame for all SNRs in both the high and low signal regime. On a GPU (CPU), the average PRUNe reconstruction times for an ensemble of interferograms is 0.013s (0.139s), whereas the SVD algorithm takes 0.303s, a speedup of ~24 (~2.2). With 32 frames per set of images in each ensemble, this theoretically corresponds to measurements speeds of up to 2432 frames-per-second for PRUNe, and 230 for SVD. We also see that in some cases, PRUNe gives sharper reconstructions than SVD. We hypothesize that this could arise when there is imperfect focusing of the phase object: whereas SVD will reconstruct the image in this defocused plane, our neural network model learns to correct the focus during training.

A drawback of our model is that it currently analyzes only 32 frames, whereas the SVD can handle a large number of frames (e.g. $10^4$ in ref. [16]). In principle, our model can also process more than 32 frames, but this imposes memory requirements on the training. In our case, the optimized models were trained with batches of 100, where each input consisted of 32 frames, using images of size 64×64 pixels, and 64 convolutional channels in the first encoder layer. Doubling any of these dimensions immediately exceeded the memory capacity (8 GB) of our GPUs, making it apparent that there are currently some limitations on the image resolution and number of frames we can process with our model. Future implementations of our model may need to include pre-processing steps that somehow combine the frames to reduce memory requirements. For example, one could consider using several principal components determined by the SVD as inputs to a NN, thereby combining the two techniques. Lastly, our model is limited in the scope of images it can recognize – here, PRUNe will likely only perform well for images of flowers or other objects with similar features, whereas the SVD algorithm is not similarly restricted.

In conclusion, we have demonstrated a 3D-2D phase retrieval U-Net model, PRUNe, that converts a 3D set of noisy, randomly phase-shifted intensity images into a single 2D output phase target image. We compared the performance of our neural network model to a state-of-the-art SVD algorithm, in the regimes of high (50 – 200 photons/pixel) and low (0.25 photons/pixel) signal, each with three different SNRs. In all cases, we find PRUNe produces smoother reconstructions with accurate phase across the entire image and corrects the image focus, whereas the SVD reconstructions can show significant pixelated noise at low SNRs and perform poorly on flat phase distributions and image backgrounds. The mean and variance of the MSE for PRUNe is consistently lower than the SVD, and this contrast is more pronounced (up to ~4-fold improvement in average MSE) in the low signal regime for few-photon images with low background counts. Our neural network approach offers a new high-throughput approach for rapid phase retrieval in intensity interferometry. Future works will consider different model architectures and the use of priors (such as the SVD reconstruction itself) that may help afford even more accurate reconstructions. While in this work we only considered spatially uniform phase jumps between frames, we would expect that with appropriate training data, our approach will also be applicable to spatially-dependent (i.e. pixelwise) phase variations due to atmospheric turbulence. Given the strong performance of our neural network model compared to established phase retrieval algorithms, we expect our 3D-2D U-Net strategy will be more broadly applicable to other imaging techniques that may require collection and combination of multiple frames that may exhibit intensity correlations obscured by noise, such as hyperspectral imaging, fluorescence lifetime or multi-photon imaging microscopy, optical coherence tomography, and photoacoustic imaging.

## Materials and methods

### Model training

The deep learning stack was implemented using PyTorch[36] and PyTorch Lightning,[37] with weight updates performed using the Adam optimizer.[38] Training was performed on NVIDIA GeForce GTX 1080 GPUs, while model evaluation and inference was performed on CPUs (2.6 GHz 6-Core Intel Core i7, 32GB RAM).

### Simulation of interferograms

Interferograms were simulated during each model training step, as a measure to prevent overfitting. First, the true phase mask is converted from grayscale to phase, and then undergoes random rotations of ±90˚ and random horizontal and/or vertical reflections. This true phase mask is used to make 32 copies, $\Phi(x, y)$. Random phase offsets, $\varphi$, between 0 and $2\pi$ are added to each copy. The simulated interferograms are generated using the below equation:[16]

$$I(x,y) = E_1^2 E_2^2 + 2V|E_1||E_2|\cos(\Phi(x,y) + \varphi))$$

Where $E_1(x,y)$ and $E_2(x,y)$ are electric fields of the interferometer laser pulses (here approximated as Gaussians), and $V$ is the visibility (for our simulations, assumed to be equal to 1). These interferograms are normalized, and then rescaled, with the addition of a flat background, to attain the target $\bar{n}_s$ and $\bar{n}_b$. A visualization of this process is giving in the SI.

### Measurement of interferograms

We describe the schematic shown in Fig. 1b in more detail. The light source is a continuous-wave laser (CrystaLaser CL532-100-S) with a center wavelength of 532 nm. Its power is attenuated to 75 uW using a series of neutral density filters. The flower phase masks are applied to the signal beam using a reflective spatial light modulator (SLM, Meadowlarks E-Series, 1920×1200). The interferometer signal and reference paths are separated by ~ 3 m on the optical table, and we observe a phase stability time of roughly 10 ms. A telescope lens ($f_0$ = 1000 mm) images the SLM plane onto the camera (UI-3240CP-NIR-GL) which has a quantum efficiency of 70%, readout noise of 88 electrons/pixel, and maximum framerate of 60 frames-per-second. A bandpass filter (Semrock LL01-532) is placed just in front of the camera to reduce background noise from room lights. We captured frames of size 256 x 256 pixels from the camera. These are then downsampled to 64 x 64 pixels to be processed by PRUNe.

### Data availability

The code used for training and testing models, generating and analyzing data, and for making figures are given in the following GitHub repository: https://github.com/andrewhproppe/PhaseRetrievalNNs. The data that supports the findings of this study are available from the corresponding authors upon reasonable request.

**Acknowledgements**

The authors thank Brayden Freitas for help with configuring the CPU and GPU devices. We acknowledge the support of the Natural Sciences and Engineering Research Council of Canada (NSERC), Canada Research Chairs, the Transformative Quantum Technologies Canada First Excellence Research Fund, and University of Ottawa-NRC Joint Centre for Extreme Quantum Photonics (JCEP) via the Quantum Sensors Challenge Program at the National Research Council of Canada.

**Conflict of interest**

The authors declare no competing interests.